\documentclass[11pt]{article}

\usepackage{amssymb}
\usepackage{epsfig}
\usepackage{amsmath}

\newtheorem{theorem}{Theorem}[section]

\newtheorem{observation}[theorem]{Observation}

\newtheorem{definition}[theorem]{Definition}

\newtheorem{lemma}[theorem]{Lemma}
\newtheorem{polyrule}{Rule}[section]
\newenvironment{proof}{\par \noindent {\bf Proof}. }
	       {\hfill$\Box$ \par \vspace{10pt}}
\def\eg{{\em e.g.}}

\def\ie{{\em i.e.}~}

\begin{document}
\title{\textbf{ Polynomial kernels for $3$-leaf power graph modification problems}\thanks{Research supported by the ANR \textsc{anr-blan-0148-06}project "\textit{Graph Decomposition on Algorithm} (GRAAL)"}}

\author{\Large \textsc{S. Bessy} \and \Large \textsc{C. Paul} \and \Large \textsc{A. Perez}}

\date{
CNRS, LIRMM, Universit\'e Montpellier 2, France~\thanks{e-mail:\tt{ bessy@lirmm.fr, paul@lirmm.fr, perez@lirmm.fr}}\\
~\\
%161, rue Ada,
%34000 Montpellier,\\
%France\\
%{\tt bessy@lirmm.fr, paul@lirmm.fr, perez@lirmm.fr}\\
\today
}

\maketitle

\begin{abstract}
  A graph $G=(V,E)$ is a $3$-leaf power iff there exists a tree $T$
  whose leaves are $V$ and such that $(u,v)\in E$ iff $u$ and $v$ are
  at distance at most $3$ in $T$. The $3$-leaf power graph edge
  modification problems, \ie edition (also known as the
  \textsc{closest $3$-leaf power}), completion and edge-deletion, are
  FTP when parameterized by the size of the edge set modification.
  However polynomial kernel was known for none of these three
  problems. For each of them, we provide cubic kernels that can be
  computed in linear time for each of these problems. We thereby
  answer an open problem first mentioned by Dom, Guo, H\"uffner and Niedermeier~\cite{DGH05}.
\end{abstract}

\newpage
%-----------------------------------------------------------------------------------------------------
%-----------------------------------------------------------------------------------------------------
\section*{Introduction}\label{S:one}

An edge modification problem aims at changing the edge set of an input
graph $G=(V,E)$ in order to get a certain property $\Pi$ satisfied
(see~\cite{NSS01} for a recent study). Edge modification problems
cover a broad range of graph optimization problems among which
completion problems (\eg ~\textsc{minimum chordal completion}),
edition problems (\eg ~\textsc{cluster edition}) and edge deletion
problems (\eg ~\textsc{maximum planar subgraph}). For completion
problems, the set $F$ of modified edges is constraint to be disjoint
from $E$, whereas for edge deletion problems $F$ has to be a subset of
$E$. If no restriction applies to $F$, then we obtain an edition
problem.  These problems are fundamental in graph theory and they play
an important role in computational theory (\eg ~they represent a large
number of the earliest NP-Complete problems~\cite{GJ79}). Edge
modification problems have recently been extensively studied in the
context of fixed parameterized complexity~\cite{DF99,Nie06}. The
natural parameterization is the number $k=|F|$ of modified edges. The
generic question is thereby whether for fixed $k$, the considered edge
modification problem is tractable. More formally:

\medskip
\noindent
\textsc{Parameterized $\Pi$-modification Problem}\\
\textbf{Input:} An undirected graph $G=(V,E)$.\\
\textbf{Parameter:} An integer $k\geqslant 0$.\\
\textbf{Question:} Is there a subset $F\subseteq V\times V$ with
$|F|\leqslant k$ such that the graph $G+F=(V,E\vartriangle F)$
satisfies $\Pi$.

\medskip This paper studies the parameterized version of edge
modification problems and more precisely the existence of a
polynomial \emph{kernel}. A problem is \emph{fixed parameterized
  tractable} (FPT for short) with respect to parameter $k$ whenever it
can be solved in time $f(k).n^{O(1)}$, where $f(k)$ is an arbitrary
function. The membership to the FPT complexity class is equivalent to
the property of having a kernel (see~\cite{Nie06} for example). A
problem is kernalizable if any instance $(G,k)$ can be reduced in
polynomial time into an instance $(G',k')$ such that the size of $G'$
is bounded by a function of $k$. Having a kernel of small size is
clearly highly desirable~\cite{GN07}. Indeed preprocessing the input
in order to reduce its size while preserving the existence of a
solution is an important issue in the context of various applications
(\cite{GN07}). However the equivalence mentioned above only provides
an exponential bound on the kernel size. For a parameterized problem
the challenge is to know whether it admits or not a polynomial, even
linear (in $k$) kernel (see \eg~\cite{Nie06}). The $k$-\textsc{vertex
  cover} problem is the classical example of a problem with a linear
kernel. Only recently, parameterized problems, among which the
\textsc{treewidth}-$k$ problem, have been shown to not have polynomial
kernel~\cite{BDF08} (unless some collapse occurs in the computational
complexity hierarchy).

In this paper we focus on graph modification problems with respect to
the property $\Pi$ being \textit{$3$-leaf power}. The $p$-power of a
graph $G=(V,E)$ is the graph $G^p=(V,E')$ with $(u,v)\in E'$ iff there
is a path of length at most $p$ between $u$ and $v$ in $G$. We say
that $G^p$ is the $p$-power of $G$ and $G$ the $p$-root of $G^p$.
Deciding whether a graph is a power of some other graphs is a well-studied problem which is NP-Complete in general~\cite{MS94}, but cubic
for $p$-power of trees (fixed $p$)~\cite{KC98}. The notion of
\emph{leaf power} has been introduced in~\cite{NRT02}: $G=(V,E)$ is a
$p$-leaf power if there exists a tree $T$ whose leaves are $V$ and
$(u,v)\in E$ iff $u$ and $v$ belong to $V$ and are at distance at most $p$ in $T$. The
$p$-leaf power recognition problem has application in the context of
phylogenetic tree reconstruction~\cite{LKJ00}. For $p\leqslant 5$, the
$p$-leaf power recognition is polynomial~\cite{KL05}. Parameterized
$p$-leaf power edge modification problems have been studied so far for
$p\leqslant 4$. The edition problem for $p=2$ is known as the
\textsc{Cluster Editing} problem for which the bound of a polynomial
kernel has been successively improved in a series of recent
papers~\cite{FLR07,GGH05,PSS07}, culminating in~\cite{Guo07} with a
$4k$ kernel size.  For larger values of $p$, the edition problem is
known as the \textsc{closest $p$-leaf power} problem. For $p=3$ and
$4$, the \textsc{closest $p$-leaf power} problem is known to be
FPT~\cite{DGH06,DGH05}, while its fixed parameterized tractability is
still open for larger values of $p$.  But the existence of a
polynomial kernel for $p\neq 2$ remained an open
question~\cite{DGH08,DGH04}. For the completion and edge-deletion, the
problems are also know to be FTP for $p\leqslant 4$~\cite{DGH08,DGH05}
and again polynomial kernel are only known for $p=2$~\cite{Guo07}.

\paragraph{\textbf{Our results.}}
We prove that the \textsc{closest $3$-leaf power}, the
\textsc{$3$-leaf power completion} and the \textsc{$3$-leaf power
  edge-deletion} admit a cubic kernel. We thereby answer positively to
the open question of Dom, Guo, H\"uffner and Niedermeier~\cite{DGH08,DGH05}.

\paragraph{\textbf{Outlines.}} 
First section is dedicated to some known and new structural results of
$3$-leaf powers and their related critical clique graphs. Section 2
describes the data-reduction rules for the \textsc{closest $3$-leaf
  power} problem. The cubic kernels for the other two variants, the
\textsc{$3$-leaf power completion} and the \textsc{$3$-leaf power
  edge-deletion} problems, are presented in Section 3.

%-----------------------------------------------------------------------------------------------------
%-----------------------------------------------------------------------------------------------------
\section{Preliminaries}

The graphs we consider in this paper are undirected and loopless
graphs. The vertex set of a graph $G$ is denoted by $V(G)$, with
$|V(G)|=n$, and its edge set by $E(G)$, with $|E(G)|=m$ (or $V$ and $E$
when the context is clear). The neighborhood of a vertex $x$ is
denoted by $N_G(x)$ (or $N(x)$). We write $d_G(u,v)$ the distance
between two vertices $u$ and $v$ in $G$. Two vertices $x$ and
$y$ of $G$ are \emph{true twins} if they are adjacent and $N(x)\cup\{x\}=N(y)\cup\{y\}$. A
subset $S$ of vertices is a \emph{module} if for any distinct vertices
$x$ and $y$ of $S$, $N(x)\setminus S=N(y)\setminus S$. Given a subset
$S$ of vertices, $G[S]$ denotes the subgraph of $G$ induced by $S$. 
If $H$ is a subgraph of $G$, $G\setminus H$ stands for $G[V(G)\setminus V(H)]$. A
graph family $\mathcal{F}$ is \emph{hereditary} if for any graph
$G\in\mathcal{F}$, any induced subgraph $H$ of $G$ also belongs to
$\mathcal{F}$. For a set $\mathcal{S}$ of graphs, we say that $G$ is
\emph{$\mathcal{S}$-free} if none of the graphs of $\mathcal{S}$ is an
induced subgraph of $G$.

As all the paper deals with undirected graphs, we abusively denote by
$X\times Y$ the set of pairs (and not couples) containing one element
of $X$ and one of $Y$.  Let $G=(V,E)$ be a graph and $F$ be a subset
of $V\times V$, we denote by $G+F$ the graph on vertex set $V$, the
edge set of which is $E\vartriangle F$ (the symmetric difference
between $E$ and $F$). Such a set $F$ is called an \emph{edition} of
$G$ (we may also abusively say that $G+F$ is an edition). A vertex
$v\in V$ is \emph{affected} by an edition $F$ whenever $F$ contains an
edge incident to $v$. Given a graph family $\mathcal{F}$ and given a
graph $G=(V,E)$, a subset $F\subseteq V\times V$ is an \emph{optimal
$\mathcal{F}$-edition} of $G$ if $F$ is a set of minimum cardinality
such that $G+F\in\mathcal{F}$. Whenever we constrain $F$ to be
disjoint from $E$, we say that $F$ is a \textit{completion}, whereas
if $F$ is asked to be a subset of $E$, then $F$ is an \textit{edge
  deletion}.

%-----------------------------------------------------------------------------------------------------
\subsection{Critical cliques}

The notions of critical clique and critical clique graph have been
introduced in~\cite{LKJ00} in the context of leaf power of graphs.
More recently, it has been successfully used in various editing
problems such as \textsc{Cluster Editing} \cite{Guo07}, \textsc{Bicluster Editing}~\cite{PSS07}.

\begin{definition}
A \emph{critical clique} of a graph $G$ is a clique $K$ which is a module and is maximal under this property.
\end{definition}

It follows from definition that the set $\mathcal{K}(G)$ of critical
cliques of a graph $G$ defines a partition of its vertex set $V$.

\begin{definition}
  Given a graph $G=(V,E)$, its \emph{critical clique graph}
  $\mathcal{C}(G)$ has vertex set $\mathcal{K}(G)$ and edge set
  $E(\mathcal{C}(G))$ with
$$(K,K')\in E(\mathcal{C}(G))\Leftrightarrow \forall v\in K, v'\in K', (v,v')\in E(G)$$
\end{definition}

Let us note that given a graph $G$, its critical clique graph
$\mathcal{C}(G)$ can be computed in linear time with modular
decomposition algorithm (see \cite{TCHP08} for example).

The following lemma was used in the construction of quadratic kernels
for \textsc{Cluster Editing} and \textsc{Bicluster Editing} problems
in~\cite{PSS07}.

\begin{lemma}\cite{PSS07} \label{lem:oneedition}
  Let $G=(V,E)$ be a graph. If $H$ is the graph
  $G+\{(u,v)\}$ with $(u,v)\in V\times V$, then
  $|\mathcal{K}(H)|\leqslant |\mathcal{K}(G)|+4$.
\end{lemma}

The following lemma shows that for special graph families, critical
cliques are robust under optimal edition.

\begin{lemma} \label{lem:closed} Let $\mathcal{F}$ be an hereditary
  graph family closed under true twin addition. For any graph
  $G=(V,E)$, there exists an optimal $\mathcal{F}$-edition (resp.
  $\mathcal{F}$-deletion, $\mathcal{F}$-completion) $F$ such that any
  critical clique of $G+F$ is the disjoint union of a subset of
  critical cliques of $G$.
\end{lemma}

\begin{proof}
  We prove the statement for the edition problem. Same arguments
  applies for edge deletion and edge completion problem.

  Let $F$ be an optimal $\mathcal{F}$-edition of $G$ such that the
  number $i$ of critical cliques of $G$ which are clique modules in
  $H=G+F$ is maximum. Assume by contradiction that $i<c$ and denote
  $\mathcal{K}(G)=\{K_1,\dots ,K_c\}$, where $K_1,\dots ,K_i$ are
  clique modules in $H$ and $K_{i+1},\dots ,K_c$ are no longer clique
  modules in $H$.  So, let $x$ be a vertex of $K_{i+1}$ such that the
  number of edges of $F$ incident to $x$ is minimum among the vertices
  of $K_{i+1}$.  Roughly speaking, we will modify $F$ by editing all
  vertices of $K_{i+1}$ like $x$.  Let $V_x$ be the subset
  $V\setminus(K_{i+1}\setminus\{x\})$.  As $\mathcal{F}$ is
  hereditary, $H_x=H[V_x]$ belongs to $\mathcal{F}$ and, as
  $\mathcal{F}$ is closed under true twin addition, reinserting
  $|K_{i+1}|-1$ true twins of $x$ in $H_x$ results in a graph $H'$
  belonging to $\mathcal{F}$.  It follows that $F'=E(G)\vartriangle
  E(H')$ is an $\mathcal{F}$-edition of $G$. By the choice of $x$, we
  have $|F'|\leqslant |F|$. Finally let us remark that by construction
  $K_1,\dots ,K_i$ and $K_{i+1}$ are clique modules of $H'$:
  contradicting the choice of $F$.
\end{proof}

In other words, for an hereditary graph family $\mathcal{F}$ which is
closed under true twin addition and for any graph $G$, there always
exists an optimal $\mathcal{F}$-edition $F$ (resp.
$\mathcal{F}$-deletion, $\mathcal{F}$-completion) such that 1) any
edge between two vertices of a same critical clique of $G$ is an edge
of $G+F$, and 2) between two distinct critical cliques
$K,K'\in\mathcal{K}(G)$, either $K\times K'\in E(G+F)$ or $(K\times
K')\cap E(G+F)=\emptyset$. From now on, every considered optimal
edition (resp. deletion, completion) is supposed to verify this
property.

%-----------------------------------------------------------------------------------------------------
\subsection{Leaf powers}

\begin{definition}
  Let $T$ be an unrooted tree whose leaves are one-to-one mapped to
  the elements of a set $V$. The \emph{$k$-leaf power} of $T$ is the
  graph  $T^k$, with $T^k=(V,E)$ where $E=\{(u,v)\mid u,v\in V
  \mbox{ and } d_T(u,v)\leqslant k\}$. We call $T$ a \emph{$k$-leaf
    root} of $T^k$.
\end{definition}

It is easy to see that for any $k$, the $k$-leaf power family of graphs satisfies Lemma~\ref{lem:closed}.
In this paper we focus on the class of $3$-leaf powers for which several characterizations is known. 

\begin{figure}
\begin{center}
\includegraphics[width=10cm]{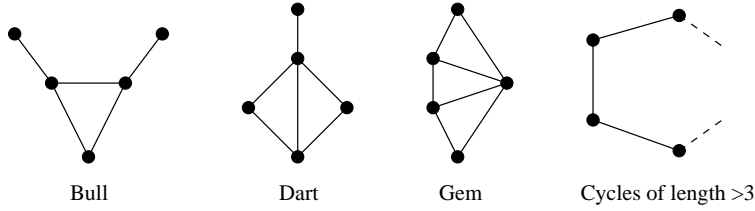}
\caption{Forbidden induced subgraphs of a $3$-leaf power.}
\label{fig:obstruction}
\end{center}
\end{figure}

\medskip
\begin{theorem}~\cite{DGH06} \label{th:carac}
For a graph $G$, the following conditions are equivalent:
\begin{enumerate}
\item $G$ is a $3$-leaf power.
\item $G$ is \{\emph{bull}, \emph{dart}, \emph{gem}, $C_{> 3}$\}-free, $C_{> 3}$ being the cycles of length at least 4. (see Figure~\ref{fig:obstruction}).
\item The critical clique graph $\mathcal{K}(G)$ is a forest.
\end{enumerate}
\end{theorem}

For the fixed parameterized
tractability of the \textsc{$3$-leaf power edition}, with respect to parameter
$k$ being the size of the editing set, the complexity
bound of $O((2k+8)^k.nm)$ is proposed in~\cite{DGH04}. The proofs of our
kernel for the \textsc{$3$-leaf power edition} problem is based on the critical
clique graph characterization and on the following new
one which uses the join composition of two graphs.

\begin{figure}[h]
\begin{center}
\includegraphics[width=8cm]{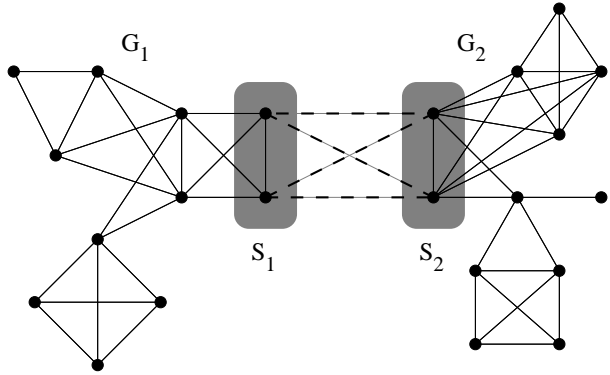}
\caption{The join composition $H=(G_1,S_1)\otimes (G_2,S_2)$ creates
  the doted edges. As $G_1$ and $G_2$ are two $3$-leaf powers and as
  the subsets $S_1$ and $S_2$ of vertices are critical cliques of
  respectively $G_1$ and $G_2$, by Theorem~\ref{th:join}, $H$ is also
  a $3$-leaf power.}
\label{fig:join}
\end{center}
\end{figure}

Let $G_1=(V_1,E_1)$ and $G_2=(V_2,E_2)$ be two disjoint graphs and let
$S_1\subseteq V_1$ and $S_2\subseteq V_2$ be two non empty subsets of
vertices. Then the \emph{join composition} of $G_1$ and $G_2$ on $S_1$
and $S_2$, denoted $(G_1,S_1)\otimes (G_2,S_2)$, results in the graph
$H=(V_1\cup V_2,E_1\cup E_2\cup (S_1\times S_2))$ (see
Figure~\ref{fig:join}).

\begin{theorem} \label{th:join} Let $G_1=(V_1,E_1)$ and
  $G_2=(V_2,E_2)$ be two connected $3$-leaf powers. Then the graph
  $H=(G_1,S_1)\otimes (G_2,S_2)$, with $S_1\subseteq V_1$ and
  $S_2\subseteq V_2$, is a $3$-leaf power if and only if one of the
  following conditions holds:
\begin{enumerate}
\item $S_1$ and $S_2$ are two cliques respectively of $G_1$ and $G_2$
  and if $S_1$ (resp. $S_2$) is not critical, then $G_1$ (resp. $G_2$)
  is a clique.
\item there exists a vertex $v\in V_1$ such that $S_1=N(v)\cup\{v\}$
  and $S_2=V_2$ is a clique.
\end{enumerate}
\end{theorem}

\begin{proof}
\begin{itemize} 
\item[$\Leftarrow$] If condition (2) holds, then we simply add true
  twins to $v$ and $H$ is a $3$-leaf power.
  Assume $S_1$ and $S_2$ are two cliques. If $S_1$ and $S_2$ are both
  critical cliques of respectively $G_1$ and $G_2$, then the critical
  clique graph $\mathcal{C}(H)$ is clearly the tree obtained from
  $\mathcal{C}(G_1)$ and $\mathcal{C}(G_2)$ by adding the edges between
  $S_1$ and $S_2$. By Theorem~\ref{th:carac}, $H$ is a $3$-leaf power.
  For $i=1$ or $2$, if $G_i$ is a clique and $S_i\subset V(G_i)$, then
  $S_i$ and $V(G_i)\setminus S_i$ are critical cliques in $H$. Again,
  it is easy to see that $\mathcal{C}(H)$ is a tree.

\item[$\Rightarrow$] First, let us notice that if $S_1$ and $S_2$ are
  not cliques, then $H$ contains a chordless $4$-cycle, which is
  forbidden. So let us assume that $S_1$ is not a clique but $S_2$ is.
  Then $S_1$ contains two non-adjacent vertices $x$ and $y$. If
  $d_{G_1}(x,y)> 2$, then $H$ contains a gem. To see this, consider
  $\pi$ a shortest $x,y$-path in $G_1$. Together with any vertex $v\in
  S_2$, the vertices of $\pi$ induce a cycle at length at least $5$ in
  $H$. By construction the only possible chords are incident to $v$.
  So any $4$ consecutive vertices on $\pi$ plus the vertex $v$ induce
  a gem. It follows that there exists in $G_1$ a vertex $u$ which
  dominates $x$ and $y$. Again, as $H$ is chordal, $u$ has to be
  adjacent to $v$ and thereby$F_2$  $u\in S_1$.  Now if there exists a
  vertex in $V(G_2)\setminus S_2$, as $G_2$ is connected, there exists
  two adjacent vertices, $v\in S_2$ and $w\in V(G_2)\setminus S_2$.
  But, $\{w,u,x,y,v\}$, induce a dart in $H$: contradicting that $H$ is
  $3$-leaf power. So, $S_2=V(G_2)$ and $G_2$ is a clique. Finally,
  assume by contraction again that $u$ has a neighbor $w\in
  V(G_1)\setminus S_1$. Considering a vertex $v$ of $S_2$, the set of
  vertices $\{w,x,y,u,v\}$ induces an obstruction in $H$, whatever
  the adjacency between $w$ and $\{x,y\}$ is. So, $N(u)\cup \{u\}\subset
  S_1$. Conversely, if $S_1$ contains a vertex $w\notin N(u)$,
  $\{w,x,y,u,v\}$ induces an obstruction in $H$. So, $S_1=N(u)\cup \{u\}$, 
  as expected in condition (2).

  Assume now that both $S_1$ and $S_2$ are cliques. If $S_1$ and $S_2$
  are not modules in respectively $G_1$ and $G_2$, then we can find a
  bull in $H$.  Assume that only $S_1$ is not a module \ie there
  exist $x,y\in S_1$ and $u\in V(G_1)\setminus S_1$ such that
  $(u,x)\in E(G_1)$ and $(u,y)\notin E(G_1)$. If $S_2\neq V(G_2)$,
  then again $H$ has a bull induced by $\{u,x,y,v,w\}$ with $v\in S_2$
  and $w\in V(G_2)\setminus S_2$, $w$ neighbor of $v$. Otherwise, either condition (2)
  holds or $y$ has a neighbor $w$ in $V(G_1)\setminus
  S_1$. The latter case is impossible since we find in
  $H$ an obstruction induced by $\{u,x,y,v,w\}$ whatever the adjacency between $w$ and
  $\{u,x\}$ is. Finally
  assume that $S_1$ and $S_2$ are modules. But consider the case that
  $S_1$ is not critical (the case $S_2$ is not critical is symmetric).
  Then there exists a critical clique $C_1\in \mathcal{K}(G_1)$
  containing $S_1$. Denote by $x$ a vertex of $S_1$ and by $y$ a vertex of $C_1\setminus
  S_1$. If $V(G_1)\neq C_1$, then $G_1$ contains two non-adjacent
  vertices, say $u$ and $u'$. If $u=x$ and $u'\notin C_1$, then as
  $G_1$ is connected, we can choose $u'$ and $w\notin C_1$ such that
  $\{u',w,x,y,v\}$ with $v\in S_2$ is a bull in $H$. Otherwise we can
  choose $u$ and $u'$ both adjacent to the vertices of $C_1$, and then
  $\{u,u',x,y,v\}$ would induce a dart in $H$. It follows that if
  $S_1$ is not critical, then condition (1) holds.
\end{itemize}
\end{proof}

In order to prove the correctness of the reduction rules, the following observation will be helpful to apply
Theorem~\ref{th:join}.

\begin{observation} \label{obs:cc} Let $C$ be a critical clique of a
  $3$-leaf power $G=(V,E)$. For any $S\subseteq V$, if $C\setminus S$
  is not a critical clique of the induced subgraph $G[V\setminus S]$,
  then the connected component of $G[V\setminus S]$ containing $C$ is
  a clique.
\end{observation}

\begin{proof}
  Assume that $C\setminus S$ is not a critical clique of $G[V\setminus
  S]$, \ie though $C\setminus S$ is a clique module in $G[V\setminus
  S]$, it is not maximal. Let $x\notin S$ be a vertex such that
  $C\cup\{x\}$ is a clique module of $G[V\setminus S]$. Then $x$
  belongs to a critical clique $C'$ of $G$ adjacent to $C$ in
  $\mathcal{C}(G)$. It follows that $S$ has to contain the union of
  all the critical cliques of $G$ adjacent to $C$ in $\mathcal{C}(G)$  but $C'$ 
 (otherwise $C\cup\{x\}$ could not be a module of $G[V\setminus S]$), and 
  all the critical cliques of $G$ adjacent to $C'$ in $\mathcal{C}(G)$  but $C$ 
 (for the same reason). 
 This means that the connected
  component containing $C$ in $G[V\setminus S]$ is a subset of $C\cup
  C'$ which is a clique.
\end{proof}

Finally, let us conclude this preliminary study of $3$-leaf powers by a technical lemma required in the proof of the last reduction rule.

\begin{lemma} \label{lem:cycle}
Let $G=(V,E)$ be a $3$-leaf power. Any cycle $C$ of length at least $5$ in $G$ contains four distinct vertices $a,b,c,d$ 
(appearing in this order along $C$) with $ab$ and $cd$ edges of $C$ such that $ad\in E$, $ac\in E$ and $bd\in E$.
\end{lemma}

\begin{proof}
As the $3$-leaf power graphs form an hereditary family, the subgraph $H$ of $G$ 
induced by the vertices of the cycle $C$ is a $3$-leaf power with at least $5$ vertices. 
As $H$ is not a tree, it contains a critical clique $K$ of size at least $2$. 
Let $a$ and $d$ be two distinct vertices of $K$. As $|C|\ge 5$, observe that there exist two distinct vertices $b$ and $c$,
distinct from $a$ and $d$,
  such that $a$, $b$, $c$ and $d$ appear in this order along $C$ and that $ab$ and $cd$ are edges of $C$. 
As $K$ is a clique module, any vertex adjacent to some vertex in $K$ neighbors all the vertices of $K$. 
It follows that $ad\in E$, $ac\in E$ and $bd\in E$.
\end{proof}

%-----------------------------------------------------------------------------------------------------
%-----------------------------------------------------------------------------------------------------
\section{A cubic kernel for the \textsc{$3$-leaf power edition} problem}

In this section, we present five preprocessing rules the application of which leads to a cubic 
kernel for the \textsc{$3$-leaf power edition} problem. The first rule is the trivial one which gets rid 
of connected components of the input graph that are already $3$-leaf powers. 
Rule~\ref{rule:trivial} is trivially safe.

\begin{polyrule} \label{rule:trivial}
If $G$ has a connected component $C$ such that $G[C]$ is  $3$-leaf power, then remove $C$ from $G$.
\end{polyrule}

The next rule was already used to obtain a quadratic kernel for the parameterized cluster editing
problem~\cite{PSS07}. It bounds the size of any critical clique in a reduced instance by $k+1$.

\begin{polyrule} \label{rule:bigCC}
If $G$ has a critical clique $K$ of size $|K|>k+1$, then remove $|K|-k-1$ vertices of $K$ from $V(G)$. 
\end{polyrule}

The safeness of Rule~\ref{rule:bigCC} follows from the fact that Lemma~\ref{lem:closed} applies to 
$3$-leaf powers. 

%-----------------------------------------------------------------------------------------------------
\subsection{Branch reduction rules}

We now assume that the input graph $G$ is reduced under Rule~\ref{rule:trivial} (\ie none of the 
connected component is a $3$-leaf power) and Rule~\ref{rule:bigCC} (\ie critical cliques of $G$ 
have size at most $k+1$). The next three reduction rules use the fact that the critical clique 
graph of a $3$-leaf power is a tree. The idea is to identify induced subgraphs of $G$, called 
\emph{branch}, which corresponds to subtrees of $\mathcal{C}(G)$. That is a branch of $G$ is an
 induced subgraph which is already a $3$-leaf power. More precisely:

\begin{definition}
Let $G=(V,E)$ be a graph. An induced subgraph $G[S]$, with $S\subseteq V$, is a \emph{branch} if
 $S$ is the disjoint union of critical cliques $K_1,\dots ,K_r\in \mathcal{K}(G)$ such that the subgraph
  of $\mathcal{C}(G)$ induced by $\{K_1,\dots ,K_r\}$ is a tree.
\end{definition}

Let $B=G[S]$ be a branch of a graph $G$ and let $K_1,\dots ,K_r$ be the critical cliques of $G$ 
contained in $S$. We say that $K_i$ ($1\leqslant i\leqslant r$) is an \emph{attachment point} of 
the branch $B$ if it contains a vertex $x$ such that $N_G(x)$ intersects $V(G)\setminus S$. A 
branch $B$ is a \emph{$l$-branch} if it has a $l$ attachment points. Our next three rules deal 
with $1$-branches and $2$-branches.

In the following, we denote by $B^R$ the subbranch of $B$ in which the vertices of the attachment
 points have been removed. If $P$ is an attachment point of $B$, then the set of neighbors of vertices 
of $P$ in $B$ is denoted $N_B(P)$.

\begin{lemma}\label{lem:1branch}
Let $G=(V,E)$ be a graph and $B$ be a $1$-branch of $G$ with attachment point $P$. There exists 
an optimal $3$-leaf power edition $F$ of $G$ such that
\begin{enumerate}
\item the set of affected vertices of $B$ is a subset of $P\cup N_B(P)$ and
\item in $G+F$, the vertices of $N_B(P)$ are all adjacent to the same vertices of $V \setminus B^R$.
\end{enumerate}
\end{lemma}
\begin{proof}
Let $F$ be an arbitrary edition of $G$ into a $3$-leaf power. We construct from $F$ a (possibly) smaller 
edition which satisfies the two conditions above.

Let $C$ be the critical clique of $H=G+F$ that contains $P$ and set $C'=C\setminus B^R$. 
By Lemma~\ref{lem:closed}, the set of critical cliques of $G$ whose vertices belong to $N_B(P)$ 
contains two kind of cliques: those, say $K_1,\dots, K_c$, whose vertices are in $C$ or adjacent 
to the vertices of $C$ in $H'$, and those, say $K_{c+1},\dots ,K_h$ whose vertices are not 
adjacent to the vertices of $C$ is $H'$. For $i\in \{1,\dots ,h\}$, let $C_i$ be the connected 
component of $B^R$ containing $K_i$.

Let us consider the three following induced subgraphs: $G_1$ the
subgraph of $G$ whose connected components are $C_1,\dots ,C_c$; $G_2$
the subgraph of $G$ whose connected components are $C_{c+1},\dots
,C_h$; and finally the subgraph $G'$ of $H$ induced by $V\setminus
B^R$.  Let us notice that these three graphs $G_1$, $G_2$ and $G'$ are
$3$-leaf power.  By Observation~\ref{obs:cc}, if $C'$ is not a
critical clique of $G'$, then the connected component of $G'$
containing $C'$ is a clique.  Similarly, if $K_i$, for any $1\leqslant
i\leqslant c$, is not a critical clique of $G_1$, the connected
component of $G_1$ in which it is contained is a clique. Thus, by
Theorem~\ref{th:join}, the disjoint union $H'$ of $G_2$ and
$(G',C')\otimes (G_1,\{K_1,\dots ,K_c\})$ is a $3$-leaf power.  Now by
construction, the edge edition set $F'$ such that $H'=G+F'$ is a
subset of $F$.  Moreover the vertices of $B$ affected by $F'$ all
belong to $P\cup N_B(P)$, which proves the first point.

  \begin{figure}[h!]
    \begin{center}
      \includegraphics[width=12cm]{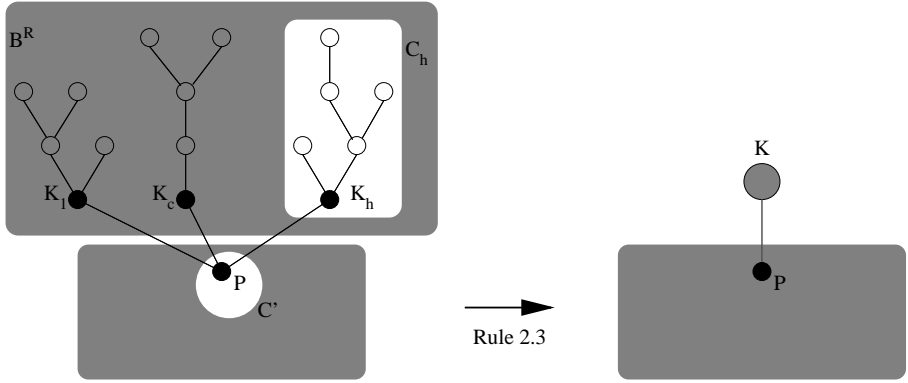} 
      \caption{On the left, a $1$-branch $B$, with attachment point
        $P$, in which the component $C_h$ of $B^R$ is distinguished in
        white. On the right, the effect of Rule~\ref{rule:1branch}
        which replace $B^R$ by a clique $K$ of size
        $\min\{|N_B(P)|,k+1\}$. }
      \label{fig:1branch}
    \end{center}
  \end{figure} 

We now consider an optimal edition $F$ that satisfies the first point. To state the second point, we focus on the 
relationship between the critical cliques $K_i$ and $C'$ in $H=G+F$.  If some $K_i$ is linked to $C'$ in $H$ (i.e. $c>1$),
 it means that the cost of adding the missing edges between $K_i$ and $C'$  (which, by Theorem~\ref{th:join}, would also 
result in a $3$-leaf power) is lower than the cost of removing the existing edge between $K_i$ and $C'$: 
$|K_i|.|C'\setminus P|\leqslant |K_i|.|P|$. On the other hand, if some $K_j$ is not linked to $C'$ in $H$ 
(i.e. $c<h$), we conclude that $|P|\leqslant |C'\setminus P|$.  Finally, if both cases occur, we have 
$|P|=|C'\setminus P|$, and we can choose to add all or none of the edges between $K_i$ and $C'$. 
In both cases, we provide an optimal edition of $G$ into a $3$-leaf powerin which, the vertices of 
$N_B(P)$ are all adjacent to the same vertices of $V\setminus B^R$.
\end{proof}

We can now state the first $1$-branch reduction rule whose safeness directly follows from Lemma~\ref{lem:1branch}.

\begin{polyrule}\label{rule:1branch}
If $G$ contains a $1$-branch $B$ with attachment point $P$, then remove from $G$ the vertices of $B^R$ and 
add a new critical clique of size $\min\{|N_B(P)|,k+1\}$ adjacent to $P$.
\end{polyrule}

Our second $1$-branch reduction rule considers the case where several $1$-branches are attached to the rest of 
the graph by a join. The following lemma shows that under certain cardinal condition, the vertices of such 
$1$-branches are not affected by an optimal edition.

\begin{lemma} \label{lem:several-1branch}
Let $G=(V,E)$ be a graph for which a $3$-leaf power edition of size at most $k$ exists.
Let $B_1,\dots ,B_l$ ($l\geqslant 2$) be $1$-branches,  the attachment points $P_1,\dots ,P_l$ 
of which all have the same neighborhood $N$ in $V\setminus \cup_{i=1}^l V(B_i)$. If $\sum_{i=1}^l |P_i|> 2k+1$, 
then, in any $3$-leaf power optimal edition $F$ of $G$, $N$ has to be a critical clique of $G+F$ and none of 
the vertices of $\cup_{i=1}^l V(B_i)$ is affected.
\end{lemma}

 \begin{proof}
We just show that any optimal $3$-leaf power edition $F$ of $G$ has to transform $N$ into a critical clique, 
which directly implies the second part of the result. First, notice that since $G$ is reduced under Rule~\ref{rule:bigCC},
 any attachment point $P_i$ satisfies $|P_i|\leqslant k+1$.

Assume that $F$ does not edit $N$ into a clique: \ie there are two vertices $a$ and $b$ of $N$ 
such that $(a,b)\notin E(G+F)$. For any pair of vertices $u_i\in P_i$ and $u_j\in P_j$ with $i\neq j$, 
the set $\{a,b,u_i,u_j\}$ cannot induce a chordless cycle in $H=G+F$, which implies that the vertices 
of $P_i$ or those of $P_j$ are affected. It follows that among the attachment points, the vertices of at 
most one are not affected by $F$. As the 
$P_i$'s have size at most $k+1$, the size of $F$ has to be at least $k+1$: contradicting the 
assumptions. So $N$ is a clique in $G+F$. 

Now, assume that $N$ is not a module of $G+F$: \ie  there exists $w\notin N$ such that for some $x,y \in N$
 we have $(x,w) \in E(G+F)$ and $(y,w) \notin E(G+F)$. As $|F|\leqslant k$, there exist two vertices 
$u_i\in P_i$ and $u_j\in P_j$, such 
that $u_iu_j\notin E(G+F)$. But, together with $x,y$ and $w$, $u_i$ and $u_j$ induce a dart in $G+F$, what 
contradicts Theorem~\ref{th:carac}. So, in $G+F$, the set of vertices 
$N$ has to be a clique module. 

Finally, let us notice that $N$ has to be critical in $G+F$, otherwise it would imply that there exists 
a vertex $v\notin N$ that has been made adjacent to at least $k+1$ vertices of $\cup_{i=1}^l B_i$, implying 
that $|F|>k$: contradiction.
\end{proof}

By Lemma~\ref{lem:several-1branch}, if there exists a $3$-leaf power edition $F$ of $G$ 
such that $|F|\leqslant k$, then the $1$-branches $B_1,\dots ,B_l$ can be safely replaced by 2 critical 
cliques of size $k+1$. This gives us the second $1$-branch reduction rule.

\begin{polyrule}\label{rule:several-1branch}
If $G$ has several $1$-branches $B_1,\dots ,B_l$ ($l\geqslant 2$), the attachment points $P_1,\dots ,P_l$ 
of which all have the same neighborhood $N$ in $V\setminus \cup_{i=1}^l V(B_i)$ and if $\sum_{i=1}^l |P_i|> 2k+1$, 
then remove from $G$ the vertices of $\cup_{i=1}^l V(B_i)$ and add two new critical cliques of size $k+1$ 
neighboring exactly $N$.
\end{polyrule}

%-----------------------------------------------------------------------------------------------------
\subsection{The $2$-branch reduction rule}

Let us consider a $2$-branch $B$ of a graph $G=(V,E)$ with attachment points $P_1$ and
$P_2$.  The subgraph of $G$ induced by the critical cliques of the unique path from 
$P_1$ to $P_2$ in $\mathcal{C}(B)$ is called the \emph{main path} of $B$ and denoted $path(B)$. 
We say that $B$ is \emph{clean} if $P_1$ and $P_2$ are leaves of $\mathcal{C}(B)$ and 
denote by $Q_1$ and $Q_2$ the critical cliques which respectively neighbor $P_1$ and $P_2$ in $B$.

\begin{lemma}\label{lem:2branch}
Let $B$ be a clean $2$-branch of a graph $G=(V,E)$ with attachment points $P_1$ and $P_2$ such that 
$path(B)$ contains at least $5$ critical cliques. Then there exists an optimal $3$-leaf power edition 
$F$ of $G$ such that
\begin{enumerate}
\item if $path(B)$ is a disconnected subgraph of $G+F$, then $F$ may contain a min-cut of $path(B)$;
\item and in any case, the other affected vertices of $B$ belongs to $P_1\cup Q_1\cup P_2\cup Q_2$.
\end{enumerate}
\end{lemma}

 \begin{proof}
Let $F$ be 
an arbitrary optimal $3$-leaf power edition of $G$. We call $C_1$ and $C_2$ the critical cliques of 
$G+F$ that respectively contain $P_1$ and $P_2$ (possibly, $C_1$ and $C_2$ could be the same), and 
denote $C_1\setminus B^R$ and $C_2\setminus B^R$ respectively by $C'_1$ and $C'_2$ 
(see Figure~\ref{fig2branch}). We will construct from $F$ another optimal $3$-leaf power 
edition $F'$ of $G$ satisfying the statement. 

  \begin{figure}[h!]
    \begin{center}
      \includegraphics[width=12cm]{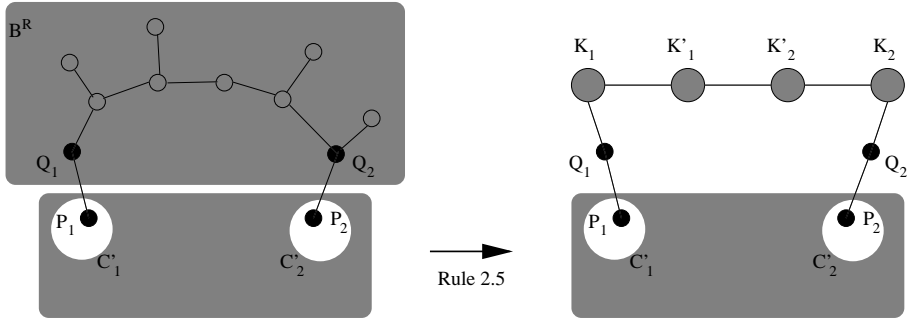}
      \caption{A $2$-branch $B$ on the left (only pendant critical
        cliques are hanging on $path(B)$ since we can assume that the
        graph is reduced by the previous rules). On the right, the way
        Rule~\ref{rule:2branch} reduces $B$.}
      \label{fig2branch}
    \end{center}
  \end{figure}

\begin{itemize}
\item \textit{Assume that $F$ disconnects $path(B)$.}
First of all, it is clear that for any subset $F_1$ of $F$, if $F_2$ is an optimal edition of $H_1=G+F_1$, then 
$F'=F_1\cup F_2$ is an optimal $3$-leaf power edition of $G$. We use this fact in the following different cases.
Assume that $F$ contains the edges $F_1:=P_1\times Q_1$ and consider the graph $H_1:=G+F_1$. We call $B_1$ the
1-branch $B\setminus P_1$ of $H_1$ whose attachment point is $P_2$. Then, Lemma~\ref{lem:1branch} applies to $B_1$
and provides from $F$ an optimal $3$-leaf power edition of $H_1$ $F_2$ where the edited vertices of $B_1$ are contained
in $P_2\cup Q_2$. By the previous observation, it follows that $F_1\cup F_2$ is an optimal edition for
$G$ that respects conditions (1) and (2). We proceed similarly if $F$ contains the edges $P_2\times Q_2$.\\
Now, we consider that $F$ does not contain $P_1\times Q_1$ or $P_2\times Q_2$. In that case, there exists $F_1\subset F$
which is a minimal cut of $path(B)$ disjoint from $P_1\times Q_1$ and $P_2\times Q_2$.
Then, there are two connected component in $B+F_1$, the one containing $P_1$, say $B_1$, and the one containing $P_2$,
say $B_2$. The subgraphs of $H_1:=G+F_1$, $B_1$ and $B_2$ are 1-branches with respectively $P_1$ and $P_2$ 
for attachment points. So, Lemma~\ref{lem:1branch} applies to $B_1$ and $B_2$ 
and provides, from $F$ an optimal $3$-leaf power edition of $H_1$ $F_2$ where the edited vertices of $B_1$ and $B_2$ 
are contained in $P_1\cup P_2\cup Q_1\cup Q_2$. To conclude, remark that, if 
$F_1$ is not a minimum (for cardinality) cut of $path(B)$, we could replace $F_1$ by such a minimum cut, and perform
a similar $3$-leaf power edition for $G$ with size strictly lower than $|F|$, what contradicts the choice of $F$.
It follows that $F_1\cup F_2$ is an optimal edition for $G$ that respects conditions (1) and (2).

\item \textit{Assume that $F$ does not disconnect $path(B)$.}
Let $X_1$ (resp. $X_2$) be the connected component of $(G+F)\setminus B^R$ containing $P_1$ (resp. $P_2$). \\
We first consider the case where $X_1$ and $X_2$ are distinct connected components. By definition, $B^R$ 
is a $3$-leaf power and $Q_1$ and $Q_2$ are two of its critical cliques (since $path(B)$ contains at least 
$5$ critical cliques). Moreover the subgraph $X_1$ (resp. $X_2$) is also a $3$-leaf power which is a clique 
if $C'_1$ (resp. $C'_2$) is not a critical clique. It follows that by Theorem~\ref{th:join}, the composition 
of these three subgraphs yields a $3$-leaf power: $H'=(X_1,C'_1)\otimes (B^R,Q_1)$ is a $3$-leaf power and 
$(H',Q_2)\otimes (X_2,C'_2)$ is a $3$-leaf power. It follows that if
$F$ affects some vertices of $V(B^R)\setminus (Q_1\cup Q_2)$, then a smaller edition could be found by removing 
from $F$ the edges in $V(B^R)\times V(B^R)$. This would contradict the optimality of $F$. 

So assume that $P_1$ and $P_2$ belongs to the same connected component $X$ of $(G+F)\setminus B^R$. Let 
$y_1$ and $y_2$ be respectively vertices of $P_1$ and $P_2$ (in the case $C_1=C_2$, choose $y_1=y_2$). Let 
$\pi_B$ and $\pi_{X}$ be two distinct paths between $y_1$ and $y_2$ defined as follows: $\pi_B$ is obtained 
by picking one vertex $b_i$ in each critical clique $H_i$ of $path(B)$ ($H_1=P_1$ and $H_q=P_2$, with 
$q\geqslant 5$); $\pi_{X}$ is a chordless path in $X$ (thereby its vertices $x_1,\dots ,x_p$, with $x_1=y_1$ and $x_p=y_2$
belong 
to distinct critical cliques, say $K_1,\dots ,K_p$ of $G+F$, with $K_1=C'_1$ and $K_p=C'_2$). 
The union of these two paths results 
in a cycle $C$ of length at least $5$. So by Lemma~\ref{lem:cycle}, there are two disjoint edges $e=(a,b)$ and $f=(c,d)$ 
in $C$ such that the edges $(a,c)$ and $(b,d)$ belong to $E \vartriangle F$. By construction of $C$, at most one of 
the edges $e$ and $f$ belongs to $\pi_{X}$.
\begin{itemize}
\item Either the edges $e$ and $f$ belong to $\pi_B$. W.l.o.g assume that $a=b_i$, $b=b_{i+1}$ and $c=b_j$, $d=b_{j+1}$ 
($i+1<j$). By Lemma~\ref{lem:closed}, $F$ contains the set of edges $(H_i\times H_j)\cup (H_{i+1}\times H_{j+1})$. 
Notice that $\min \{|H_i|.|H_{i+1}|, |H_j|.|H_{j+1}|\} < |H_i|.|H_j|+|H_{i+1}|.|H_{j+1}|$. W.lo.g. assume that 
$\min \{|H_i|.|H_{i+1}|, |H_j|.|H_{j+1}|\}=|H_i|.|H_{i+1}|$. We will 'cut' the edges between $H_i$ and $H_{i+1}$:
consider the set 
$$F'=(F\setminus (V\times V(B^R)))\cup (H_i\times H_{i+1})$$ 
Moreover, if $H_i\neq P_1$, add to $F'$ the edges $(C'_1\setminus P_1)\times Q_1$ (which were previously in $F$)
and, if $H_{i+1}\neq P_2$, add to $F'$ the edges $(C'_2\setminus P_2)\times Q_2$ (which were previously in $F$).
In all cases, we have $|F'|<|F|$. As in the case where $X_1$ and $X_2$ were distinct, by Theorem~\ref{th:join}, 
the graph $G+F'$ is a $3$-leaf power: contradicting the optimality of $F$.

\item Or the edge $e$ belongs to $\pi_B$ and $f$ to $\pi_X$. W.l.o.g. assume that $a=b_i\in H_i$, $b=b_{i+1}\in H_{i+1}$ 
and $d=k_j\in K_j$, $c=k_{j+1}\in K_{j+1}$. As above, by Lemma~\ref{lem:closed}, $F$ contains 
$(H_i\times K_{j+1})\cup (H_{i+1}\times K_j)$.
Notice that $\min \{|H_i|.|H_{i+1}|, |K_j|.|K_{j+1}|\} < |H_i|.|K_{j+1}|+|H_{i+1}|.|K_j|$.
If $\min \{|H_i|.|H_{i+1}|, |K_j|.|K_{j+1}|\}=|H_i|.|H_{i+1}|$, 
then we consider the set  
$$F'=(F\setminus (V\times V(B^R)))\cup (H_i\times H_{i+1})$$
Here again, if $H_i\neq P_1$, add to $F'$ the edges $(C'_1\setminus P_1)\times Q_1$ (which were previously in $F$)
and, if $H_{i+1}\neq P_2$, add to $F'$ the edges $(C'_2\setminus P_2)\times Q_2$ (which were previously in $F$).
As previously, $|F'|$ is smaller than $|F|$ and  by Theorem~\ref{th:join}, we can prove that $G+F'$ is a $3$-leaf power.
Finally, if $\min \{|H_i|.|H_{i+1}|, |K_j|.|K_{j+1}|\}=|K_j|.|K_{j+1}|$, then we consider the set

\smallskip
\noindent
$F'=(F\setminus (V\times V(B^R)))\cup (K_i\times K_{i+1})\cup ((C'_1\setminus P_1)\times Q_1)\cup ((C'_2\setminus P_2)\times
 Q_2)$

\smallskip
\noindent 
Again $|F'|$ is smaller than $|F|$ and  by Theorem~\ref{th:join}, we can prove that $G+F'$ is a $3$-leaf power. 
In any case, we found a better $3$-leaf power edition $F'$, contradicting the optimality of $F$.
\end{itemize}
\end{itemize}
\end{proof}

\begin{polyrule}\label{rule:2branch}
Let $G$ be a graph having a clean $2$-branch $B$ such that $path(B)$ is composed by at least $8$ critical cliques.
Then remove from $G$ all the vertices of $V(B)$ except those of $P_1\cup Q_1\cup P_2\cup Q_2$ and add four new 
critical cliques:
\begin{itemize}
\item $K_1$ (resp. $K_2$) of size $k+1$ adjacent to $Q_1$ (resp. $Q_2$);
\item $K'_1$ (resp $K'_2$) adjacent to $K_1$ (resp. $K_2$) and such that $K'_1$ and $K'_2$ are adjacent and $|K'_1|.|K'_2|$ equals the min-cut of $path(B)$.
\end{itemize}
\end{polyrule}

\begin{proof}
  Let $B'$ be the $2$-branch replacing $B$ after the application of
  the rule. It is easy to see that by construction the min-cut of $B'$
  equals the min-cut of $path(B)$. Moreover the attachment points $P_1$
  and $P_2$ and their respective neighbors $Q_1$ and $Q_2$ are
  unchanged. It follows by Lemma~\ref{lem:2branch} that any optimal
  edition $F$ of $G$ corresponds to an optimal edition $F'$ of $G'$,
  the graph reduced by Rule~\ref{rule:2branch}, such that $|F|=|F'|$.
\end{proof}

%-----------------------------------------------------------------------------------------------------
\subsection{Kernel size and time complexity}

Let us discuss the time complexity of the reduction rules. The
$3$-leaf power recognition problem can be solved in
$O(n+m)$~\cite{BL06}. It follows that Rule~\ref{rule:trivial} requires
linear time. To implement the other reduction rules, we fist need to
compute the critical clique graph $\mathcal{C}(G)$. As noticed
in~\cite{PSS07}, $\mathcal{C}(G)$ can be built in $O(n+m)$. For
instance, to do so, we can compute in linear time the modular
decomposition tree of $G$, which is a classical and well-studied
problem in algorithmic graph theory (see~\cite{TCHP08} for a recent
paper). Given $\mathcal{K}(G)$, which is linear in the size of $G$, it
is easy to detect the critical cliques of size at least $k+1$. So,
Rule~\ref{rule:bigCC} requires linear time. A search on
$\mathcal{C}(G)$ can identify the $1$-branches.  It follows that the
two $1$-branches reduction rules (Rule~\ref{rule:1branch} and
Rule~\ref{rule:several-1branch}) can also be applied in $O(n+m)$ time.
Let us now notice that in a graph reduced by the first four reduction
rules, a $2$-branch is a path to which pendant vertices are possibly
attached. It follows that to detect a $2$-branch $B$, such that
$path(B)$ contains at least $5$ critical cliques, we first prune the
pendant vertices and then, identify in $\mathcal{C}(G)$ the paths
containing at least 5 vertices (for instance, by proceeding a DFS
starting on $\mathcal{C}(G)$ at a vertex of degree at least $3$, if it
exists, otherwise the problem is trivial).  This shows that
Rule~\ref{rule:2branch} can be carried in linear time.

\begin{theorem} \label{th:kernel}
The parameterized \textsc{$3$-leaf power edition} problem admits a cubic kernel. Given a graph $G$, a reduced instance can be 
computed in linear time.
\end{theorem}

\begin{proof}
The discussion above established the time complexity to compute a kernel. Let us  determine the kernel size.
Let $G=(V,E)$ be a reduced graph (\ie none of the reduction rules applies to $G$) which can be edited into a 
$3$-leaf power with a set $F\subseteq V\times V$ such that $|F|\leqslant k$. Let us denote $H=G+F$ the edited graph. 
We first show that $\mathcal{C}(H)$ has $O(k^2)$ vertices (\ie $|\mathcal{K}(H)|\in O(k^2)$). Then 
Lemma~\ref{lem:oneedition} enables us to conclude.

We say that a critical clique is affected if it contains an affected vertex and denote by $A$ the set of the 
affected critical cliques. As each edge of $F$ affects two vertices, we have that $|A|\leqslant 2k$.  Since 
$H$ is a $3$-leaf power, its critical clique graph $\mathcal{C}(H)$ is a tree. Let $T$ be the minimal subtree 
of $\mathcal{C}(H)$ that spans the affected critical cliques. Let us observe that if $B$ is a maximal subtree of 
$\mathcal{C}(H)-T$, then none of the critical cliques in $B$ contains an affected vertices and thus $B$ was the 
critical clique graph of a $1$-branch of $G$, which has been reduced by Rule~\ref{rule:1branch} or 
Rule~\ref{rule:several-1branch}. Let $A'\subset \mathcal{K}(H)$ be the critical cliques of degree at least 
$3$ in $T$. As $|A|\leqslant 2k$, we also have $|A'|\leqslant 2k$. The connected components resulting from 
the removal of $A$ and $A'$ in $T$ are paths. There are at most $4k$ such paths. Each of these paths is 
composed by non-affected critical cliques. 
It follows that each of them corresponds to $path(B)$ for some $2$-branch $B$ of $G$, which has been reduced by 
Rule~\ref{rule:2branch}. From these observations, we can now estimate the size of the reduced graph. 
Attached to each of the critical cliques of $T\setminus A$, we can have $1$ pendant critical clique 
resulting from the application of Rule~\ref{rule:1branch}. Remark that any 2-branch reduced by Rule~\ref{rule:2branch}
has no such pendant clique and that $path(B)$ contains at least $8$ critical cliques. So, a considered 2-branch in 
$\mathcal{C}(H)$ is made of at most 14 critical cliques.
Finally attached to each critical clique of $A$, we can have at most $(4k+2)$ extra critical 
cliques resulting from the application of Rule~\ref{rule:several-1branch}. See Figure~\ref{fig:size} for an 
illustration of the shape of $\mathcal{C}(H)$. Summing up everything, we obtain that 
$\mathcal{K}(H)$ contains at most $4k.14+2k.2+2k.(4k+3)=8k^2+66k$ critical cliques.

\begin{figure}[h]
\begin{center}
\includegraphics[width=12cm]{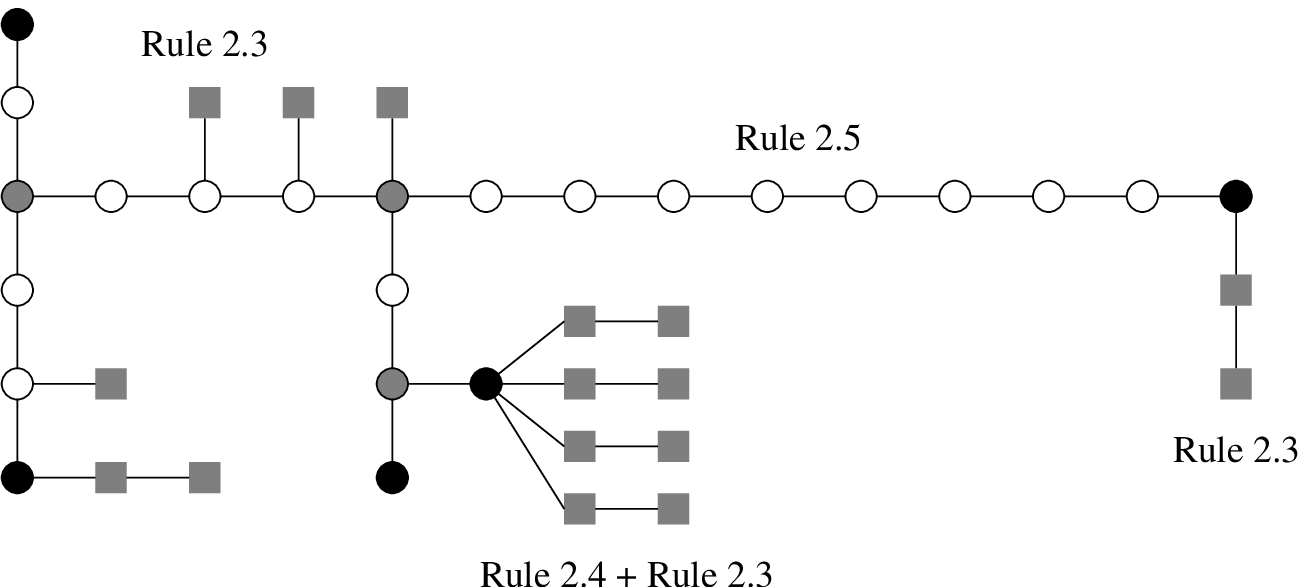}
\caption{The black circles are the critical cliques of $A$, the grey
  ones belong to $A'$, and the squares are the critical cliques not in
  $T$. On the figure, we can observe a $2$-branch of size $8$ reduced
  by Rule~\ref{rule:2branch}.  There cannot be pendant critical
  cliques attached to its nodes. Application of
  Rule~\ref{rule:1branch}, may let a path of two critical cliques
  pendant to the elements of $A\cup A'$ and a single critical clique
  pendant to the elements of the small $2$-branches. Finally,
  Rule~\ref{rule:several-1branch} can only affect critical cliques of
  $A$.}
\label{fig:size}
\end{center}
\end{figure}

By Lemma \ref{lem:oneedition}, we know that each edited edge in a graph, the number of
critical cliques increase of  at most 4. It follows that $\mathcal{K}(G)$ contains at most $8k^2+70k$ critical cliques. 
By Rule~\ref{rule:bigCC}, each critical clique of the reduced graph has size at most $k+1$. This implies that the 
reduced graph contains at most $8k^3+78k^2+70k$ vertices, proving the $O(k^3)$ kernel size.
\end{proof}

We should notice that some small modifications of the branch reduction rules and a more precise analysis would 
improve the constants involved in the kernel size. However the cubic bound would not change.

%-----------------------------------------------------------------------------------------------------
%-----------------------------------------------------------------------------------------------------
\section{Cubic kernels for edge completion only and edge deletion only}

We now prove and/or adapt the previous rules to the cases where only
insertions or only deletions of edges are allowed. First, observe that
Rules \ref{rule:trivial} and \ref{rule:bigCC} are also safe in \textsc{3-leaf
power completion} and \textsc{$3$-leaf power deletion} (Rule \ref{rule:bigCC}
directly follows from Lemma \ref{lem:closed}). We have a similar
result for the $1$-branches reduction rules.

\begin{lemma} \label{lemma:both1branch}
Rule \ref{rule:1branch} is safe for both \textsc{$3$-leaf power completion} and \textsc{$3$-leaf power deletion}.
\end{lemma}

\begin{proof}
In the following, we consider an optimal solution $F$ such that $H := G+F$
is a $3$-leaf power, denote by $C$ the critical clique containing
$P$ in $H$ and set $C' = C \setminus B^R$.

\begin{itemize}
\item \textit{\textsc{$3$-leaf power completion.}}  To show the safeness of Rule
  \ref{rule:1branch} in this case, we will build from $F$ an optimal
  completion that respects conditions of Lemma \ref{lem:1branch}. 
  By Lemma \ref{lem:closed}, we know that the set of
  critical cliques $\{K_1,\dots,K_h\}$ of $G$ whose vertices belong to
  $N_B(P)$ are in $C$ or adjacent to the vertices of $C$ in $H$ (in
  this case, there is no critical cliques $K_i$ disconnected from $C$
  in $H$ because we cannot remove edges from $G$). 
  In both cases, $K_i$ is adjacent to $C'$ in $H$. For $i \in
  \{1,\dots,h\}$, let $C_i$ be the connected component of $B^R$
  containing $K_i$.
  As previously, we consider $G_1$ the subgraph of $G$ whose connected components
  are $C_1,\dots,C_h$. By Observation \ref{obs:cc}, if $C'$ is not a
  critical clique of $G'$, then $G'$ is a clique. Similarly, if $K_i$,
  for any $1 \leq i \leq h$, is not a critical clique of $G_1$, the
  connected component of $G_1$ in which it is contained is a clique.
  By Theorem \ref{th:join}, it follows that the graph $H' := (H\setminus B^R,C')
  \otimes (G_1,\{K_1,\dots,K_h\})$ is a $3$-leaf power. By construction,
  the edge completion set $F'$ such that $H' = G + F'$ is a subset of $F$
  and the vertices of $B$ affected by $F$ all belong to $P \cup
  N_B(P)$. Finally, as every $K_i$ is connected to $C'$ in $H'$, the
  vertices of $N_B(P)$ are all adjacent to the same vertices of $V
  \setminus B^R$.

\item \textit{\textsc{$3$-leaf power deletion.}}  In the case where only edges
  deletion are allowed, we will build from $F$ an optimal deletion
  respecting the conditions of Lemma \ref{lem:1branch} by studying the
  behavior of $P$ in $H$. First of all, note that if $P$ forms a
  bigger critical clique in $H$ with some vertex $x \in V \setminus
  B^R$, this means that $F$ contains $P \times N_B(P)$. Thus, there
  is no need to do extra deletions in  $B^R$ and then we are done.

  Now consider the cases where $P$ is critical in $H$ or form a bigger
  critical clique with some $K_i$ (\emph{i.e.} $F$ contains $P \times
  (\{K_1,\dots,K_c\} \setminus K_i$ for some $i$). In both cases, we have $C' =
  P$. By Theorem \ref{th:join}, the graph $H' := (H\setminus B^R,C') \otimes (G_1,
  \{K_1,\dots,K_c\})$ is a $3$-leaf power, and the edge set $F'$ used
  to transform $G$ into $H'$ is a subset of $F$ (all the edges between
  $C'$ and $\{K_1,\dots,K_c\})$ are present in $H$), and then we are
  done.

\end{itemize}
\end{proof}

\begin{lemma} \label{lemma:bothseveral-1branch}
Rule \ref{rule:several-1branch} is safe for both \textsc{$3$-leaf power completion} and \textsc{$3$-leaf power deletion}.
\end{lemma}

\begin{proof}
As in Lemma~\ref{lem:several-1branch}, we consider $B_1,\dots ,B_l$ 1-branches of $G$,
the attachment points $P_1,\dots ,P_l$ of which all have the same neighborhood
$N$ and satisfy $\sum_{i=1}^l |P_i|> 2k+1$.

\begin{itemize}
\item{\textit{\textsc{$3$-leaf power completion.}}}  In this case, same arguments
  as the ones used in the proof of Lemma \ref{lem:several-1branch}
  hold. We briefly detail them. First, assume that $N$ was not
  transformed into a clique by an optimal completion $F$. To get rid
  of all the $C_4$'s involving 2 non-adjacent vertices of $N$ and
  $P_i$, $P_j$, $i \neq j$, the only possibility is to transform
  $\textstyle{ \cup_{i=1}^l} P_i$ into a clique, which requires more
  than $k+1$ completions. Moreover, $N$ must also become a clique
  module, otherwise we would find darts that once again would imply to
  transform $\textstyle{ \cup_{i=1}^l} P_i$ into a clique which is
  impossible. Finally, $N$ must be critical (otherwise, at least one
  insertion for each vertex of $\textstyle{ \cup_{i=1}^l} P_i$ must be
  done), thus implying that no vertex in $\textstyle{ \cup_{i=1}^l}
  P_i$ is affected by an optimal edition.
\item{\textit{\textsc{$3$-leaf power deletion.}}}  Firstly, observe that if $N$
  is not a clique, then any optimal deletion in that case would have to
  destroy at least $k+1$ $C_4$ with edges deletion only, which is
  impossible. The same arguments used previously hold again in this case to
conclude that $N$ must become a critical clique in the modified graph.
\end{itemize}
\end{proof}

Now, observe that the 2-branch reduction rule can apply directly to
\textsc{$3$-leaf power deletion}, but will not be safe for $3$-leaf power
completion. Indeed, in the proof of Lemma~\ref{lem:2branch}, if we
look at the cycle $C$ of $G$ containing vertices of $B$, it might be
needed to delete edges between two consecutive critical cliques along
$C$.  to transform $\mathcal{C}(C)$ into a tree.  Nevertheless, it is
possible to bound the number of vertices of $path(B)$ in the case of
\textsc{$3$-leaf power completion} by looking at the edges modifications needed
to make a cycle chordal (see Lemma \ref{lem:add2branch}).

\begin{lemma} \label{lem:del2branch}
Rule \ref{rule:2branch} is safe for \textsc{$3$-leaf power deletion}.
\end{lemma} 

\begin{proof}
  Let $F$ be an arbitrary optimal $3$-leaf power deletion of $G$. We
  call $C_1$ and $C_2$ the critical cliques of $H := G +F$ that
  respectively contain $P_1$ and $P_2$, $C'_1 := C_1 \setminus B^R$
  and $C'_2 := C_2 \setminus B^R$. We will construct from $F$
  another optimal $3$-leaf power edition $F'$ of $G$ satisfying the
  conditions of Lemma \ref{lem:2branch}. \par
  We have two cases to consider : $1)$ either $path(B)$ is
  disconnected in $H$ or $2)$ $path(B)$ is still connected in $H$.
  Case $1)$ works exactly as the first case studied in the
  proof of Lemma \ref{lem:2branch}, and thus there exists an optimal
  deletion on which conditions of Lemma \ref{lem:2branch} holds. \par
  If case $(2)$ holds, \emph{i.e.} if $path(B)$ is still connected in
  $H$, then $P_1$ and $P_2$ must belong to distinct connected
  components of $H\setminus B^R$, say $X_1$ and $X_2$
  (otherwise $H$ would admit a chordless cycle as induced subgraph).
  Furthermore, notice that we must have $P_1 = C_1$ and $P_2 = C_2$ in
  $H$. Indeed, if $P_1$ forms a critical clique with some vertex $x
  \in V \setminus B^R$, this means $F$ must contain $P_1 \times Q_1$
  which is not, by hypothesis. Similarly, if $P_1$ forms a critical
  clique with some vertex $x \in Q_1$, then $F$ must contain edges
  between $Q_1$ and $N_{B^R}(Q_1)$ which is not (the cases for $P_2$ are
  symmetric). By definition, $B^R$ is a $3$-leaf power, and so are
  $X_1$ and $X_2$. By Theorem \ref{th:join}, it follows that the
  composition of these three subgraphs yields a $3$-leaf power : $H' =
  (X_1,P_1) \otimes (B^R, Q_1)$ and $(H',Q_2) \otimes (X_2,P_2)$ are
  $3$-leaf powers. It follows that if $F$ affects some vertices of
  $B^R \setminus (Q_1 \cup Q_2)$, then a smaller deletion could
  be found, what contradicts the optimality of $F$.
\end{proof}

We now prove a result usefull to conclude on the size of the kernel in
the \textsc{$3$-leaf power completion} problem.

\begin{lemma} \label{lem:add2branch} Let $G$ be a graph admitting a
  clean $2$-branch $B$ such that $path(B)$ is composed by at least
  $k+4$ critical cliques. If $P_1$ and $P_2$ belong to
  the same connected component in $G$, then there is no $3$-leaf power
  completion of size at most $k$.
\end{lemma}

\begin{proof}
  Let $G$ be a graph with a clean $2$-branch $B$ on which conditions
  of the Lemma \ref{lem:add2branch} applies, and let $F$ be an optimal
  $3$-leaf power completion of $G$. As $P_1$ and $P_2$ belong to the
  same connected component in $G$, we have a cycle $C$ of size at
  least $k+4$ in $\mathcal{K}(G)$. Consider the subgraph of
  $\mathcal{K}(G)$ induced by the critical cliques of $C$. By Lemma
  \ref{lem:closed} we know that $\mathcal{C}(C)$ must be a tree ; let
  us call $F'$ the set of edges transforming $C$ into a tree. It is
  known that $F'$ is a triangulation of this cycle~\cite{DGH04}.
  Moreover, every
  triangulation of a $n$-cycle needs at least $n-3$ chords, what implies that
  $|F'| > k$, which is impossible.
\end{proof}

\begin{polyrule} \label{rule:add2branch} Let $G$ be a graph having a
  clean $2$-branch $B$ with attachment points $P_1$ and $P_2$ such
  that $path(B)$ is composed by at least $k+3$ critical cliques.
\begin{itemize} 
\item if $P_1$ and $P_2$ belong to the same connected component in
  $G\setminus B^R$, then there is no completion of size at most $k$.
\item otherwise, remove from $G$ all the vertices of $V(B)$ except
  those of $P_1 \cup Q_1 \cup P_2 \cup Q_2$ and add all possible edges
  between $Q_1$ and $Q_2$.
\end{itemize}
\end{polyrule}

\begin{proof}
  The first point follows directly from Lemma \ref{lem:add2branch}. To
  see the second point, notice that we are in the case where $P_1$ and
  $P_2$ belong to different connected components (which corresponds to
  the second case of the proof of Lemma \ref{lem:2branch}). As edges
  insertion are allowed, the safeness of this rule is due to this
  particular case.
\end{proof}

\begin{theorem}
The parameterized \textsc{$3$-leaf power completion} and \textsc{$3$-leaf power deletion} problem admit cubic kernels. Given a graph $G$ a reduced instance can be computed in linear time.
\end{theorem}

\begin{proof}
We detail separately completion and deletion.
\begin{itemize}
\item {\textit{\textsc{$3$-leaf power completion.}}}  As in the proof of Theorem
  \ref{th:kernel}, we consider $H := G+F$ with $G$ being reduced and
  $F$ being an optimal completion and we denote by $T$ the minimal
  subtree of $\mathcal{C}(H)$ spanning the set of affected critical
  cliques $A$. As noticed before, we have $|A| \leqslant 2k$. \par
  First, remark that the only difference between this case and
  \textsc{$3$-leaf power edition} concerns the $2$-branch reduction rule. This
  means that the only difference will occur in the number of vertices
  of the paths resulting from the removal of $A$ and $A'$ in $T$ ($A'$
  being critical cliques of degree at least $3$ in $T$). Due to both
  Lemma \ref{lem:add2branch} and Rule \ref{rule:add2branch} we know
  that a $2$-branch in $\mathcal{C}(H)$ is made of at most $2k + 6$
  critical cliques: corresponding to a path of at most $k+3$ critical
  cliques (otherwise there is no optimal completion), each one having
  a pendant critical clique (by Rule \ref{rule:1branch}). This means
  that $\mathcal{C}(H)$ contains at most $4k.(2k+6)+2k.2+2k.(4k+3) =
  16k^2 + 34k$ critical cliques.  By Lemma \ref{lem:oneedition}, we
  know that each edited edge creates at most $4$ new critical cliques.
  If follows that $\mathcal{K}(G)$ contains at most $16k^2 + 38k$
  critical cliques. By Rule \ref{rule:bigCC}, each critical clique of
  the reduced graph has size at most $k+1$, thus implying that the
  reduced graph contains at most $16k^3 + 54k^2 + 38k$ vertices,
  proving the $O(k^3)$ kernel size.
\item {\textit{\textsc{$3$-leaf power deletion.}}} The rules used for the 3-leaf
  power deletion problem are exactly the same than the one used to
  obtain a cubic kernel for \textsc{$3$-leaf power edition}. Thus, the size of
  a reduced instance of $3$-leaf power deletion will be exactly the
  same as one of a reduced instance of \textsc{$3$-leaf power edition}.
\end{itemize}
\end{proof}

\section{Conclusion}

By proving the existence of a cubic kernel of the \textsc{
  $3$-leaf power edition} problem, we positively answered an open
problem~\cite{DGH08,DGH05}. The natural question is now whether the
cubic bound could be improved. A strategy could be, as for the
quadratic kernel of \textsc{$3$-hitting set}~\cite{NRT04} which is
based of the linear kernel of \textsc{vertex cover}~\cite{Nie06}, to
consider the following subproblem:

\medskip
\noindent
\textsc{parameterized fat star edition problem}\\
\textbf{Input:} An undirected graph $G=(V,E)$.\\
\textbf{Parameter:} An integer $k\geqslant 0$.\\
\textbf{Question:} Is there a subset $F\subseteq V\times V$ with
$|F|\leqslant k$ such that the graph $G+F=(V,E\vartriangle F)$
is a $3$-leaf power and its critical clique graph $\mathcal{C}(G+F)$
is a star (we say that $G+F$ is a \textit{fat star}).

\medskip It can be shown that small modifications of the
Rule~\ref{rule:trivial}, \ref{rule:bigCC} and \ref{rule:several-1branch} yield a
quadratic kernel for the \textsc{fat star edition}
problem~\cite{Per08}. A linear bound may be helpful to improve the
kernel of the \textsc{closest $3$-leaf power} since it can be shown
that the neighborhood of any big enough critical clique of the input graph as to
be edited into a fat star.

%-----------------------------------------------------------------------------------------------------
%-----------------------------------------------------------------------------------------------------

\bibliographystyle{plain}
\bibliography{tech-report-3leaf}

%\end{thebibliography}

\end{document}